\documentclass{jfm}

\usepackage{graphicx}
\usepackage{natbib}

\ifCUPmtlplainloaded \else
 \checkfont{eurm10}
 \iffontfound
   \IfFileExists{upmath.sty}
     {\typeout{^^JFound AMS Euler Roman fonts on the system,
                  using the 'upmath' package.^^J}%
      \usepackage{upmath}}
     {\typeout{^^JFound AMS Euler Roman fonts on the system, but you
                  dont seem to have the}%
      \typeout{'upmath' package installed. JFM.cls can take advantage
                of these fonts,^^Jif you use 'upmath' package.^^J}%
     }
 \else
 \fi
\fi

\ifCUPmtlplainloaded \else
 \checkfont{msam10}
 \iffontfound
   \IfFileExists{amssymb.sty}
     {\typeout{^^JFound AMS Symbol fonts on the system, using the
               'amssymb' package.^^J}%
      \usepackage{amssymb}%
        \let\leq=\leqslant
        
     }{}
 \fi
\fi

\ifCUPmtlplainloaded \else
 \IfFileExists{amsbsy.sty}
   {\typeout{^^JFound the 'amsbsy' package on the system, using it.^^J}%
    \usepackage{amsbsy}}
   {\providecommand\boldsymbol[1]{\mbox{\boldmath $##1$}}}
\fi

\providecommand\bnabla{\boldsymbol{\nabla}}
\providecommand\bcdot{\boldsymbol{\cdot}}

\newsavebox{\astrutbox}
\sbox{\astrutbox}{\rule[-5pt]{0pt}{20pt}}

\newcommand\thalf{\ensuremath{{\textstyle\frac{1}{2}}}}

\newcommand\Xsin{\mbox{sin}}
\newcommand\Xcos{\mbox{cos}}
\newcommand\Xsec{\mbox{sec}}
\newcommand\Xlog{\mbox{ln}}

\title[Anisotropic electro-osmotic flow over super-hydrophobic surfaces]{Anisotropic electro-osmotic flow over super-hydrophobic surfaces}

\author[S. S. Bahga,  O. I. Vinogradova, and M. Z. Bazant ]%
{S\ls U\ls P\ls R\ls E\ls E\ls T\ns S.\ns B\ls A\ls H\ls G\ls A,$^1$%
\ns
O\ls L\ls G\ls A\ns I.\ns V\ls I\ls N\ls O\ls G\ls R\ls A\ls D\ls O\ls V\ls A,$^2$\break
\and 
M\ls A\ls R\ls T\ls I\ls N\ns Z.\ns B\ls A\ls Z\ls A\ls N\ls T$^{1,3}$
}

\affiliation{$^1$Department of Mechanical Engineering, Stanford University,
Stanford, CA 94305, USA\\[\affilskip]
$^2$ A. N. Frumkin Institute of Physical Chemistry and Electrochemistry, Russian Academy of Sciences,
31 Leninsky Prospect, 119991 Moscow, Russia \\[\affilskip]
$^3$ Departments of Chemical Engineering and Mathematics, Massachusetts Institute of Technology, Cambridge, MA~02139, USA
}
 \date{\today}
\label{firstpage}

\begin{document}

\maketitle

\begin{abstract}
Patterned surfaces with large effective slip lengths, such as super-hydrophobic surfaces containing trapped gas bubbles, have the potential to greatly enhance electrokinetic phenomena. Existing theories assume either homogeneous flat surfaces or patterned surfaces with thin double layers (compared to the texture correlation length) and thus predict simple surface-averaged, isotropic flows (independent of orientation).  By analyzing electro-osmotic flows over striped slip-stick surfaces with arbitrary double-layer thickness, we show that surface anisotropy generally leads to a tensorial electro-osmotic mobility and subtle, nonlinear averaging of surface properties. Interestingly, the electro-osmotic mobility tensor  is not simply related to the hydrodynamic slip tensor, except in special cases. Our results imply that significantly enhanced electro-osmotic flows over super-hydrophobic surfaces are possible, but only with charged liquid-gas interfaces.
\end{abstract}

\section{Introduction}
The development of microfluidics has motivated interest in 
manipulating flows in very small channels, which exhibit huge hydrodynamic
resistance to pressure-driven flow~\citep{stone2004,squires2005}. One avenue for driving flow on such scales is to exploit hydrodynamic slip, usually quantified by the slip
length $b$ (the distance within the solid at which the flow
profile extrapolates to zero)~\citep{vinogradova1999,lauga2005,bocquet2007}. For hydrophobic smooth and homogeneous surfaces $b$ can be of the order of tens of nanometers ~\citep{vinogradova2003,charlaix.e:2005,joly.l:2006,vinogradova.oi:2009}, but not much more. Since the efficiency of hydrodynamic slippage is determined by the ratio of $b$ to the scale of the channel $h$~\citep{vinogradova.oi:1995a}, it is impossible to benefit of such a nanometric slip for pressure-driven microfluidic applications. 

In principle, this limitation does not apply to interfacially-driven flows, such as electro-osmosis past a charged surface in response to an applied electric field.  The combination of these two strategies can yield considerably enhanced electro-osmotic (EO) flow on hydrophobic surfaces~\citep{muller.vm:1986,joly2004,ajdari.a:2006}, even for nanometric slip lengths. The reason is that the thickness of the electrical Debye layer (EDL), characterized by the Debye screening length $\lambda_D=\kappa^{-1}$, defines an additional length scale of the problem, comparable to $b$.  For a small surface charge density $q$, simple arguments show that the electro-osmotic mobility $M_e$, which relates the effective electro-osmotic slip velocity (outside the double layer) to the tangential electric field $u_s =M_e E_t$, is given by ~\citep{muller.vm:1986,joly2004}:
\begin{equation}
M_e = - \frac{\varepsilon \zeta}{\eta}(1+b\kappa) = - \frac{q}{\eta\kappa}(1+b\kappa),    \label{isotropic}
\end{equation}
where $\varepsilon$ and $\eta$ are the permittivity and viscosity of the solution, respectively, and $\zeta=q/\kappa\varepsilon$ is the zeta potential across the diffuse (flowing) part of the double layer.  The factor $(1+b\kappa)$ associated with hydrodynamic slip can potentially enhance interfacially-driven flow in microfluidic devices~\citep{ajdari.a:2006}, electrophoretic mobility of particles~\citep{khair.as:2009}, and electrokinetic energy conversion (streaming potential) in  nanochannels~\citep{heyden2006}. 

For this reason, it is attractive to consider electro-osmotic flows over superhydrophobic (SH) surfaces, whose texture on  a scale $L$  can significantly amplify hydrodynamic slip due to gas entrapment~\citep{vinogradova.oi:1995b,cottin_bizonne.c:2003.a} leading to effective $b$ of the order of several microns in pressure driven flows~\citep{ou2005,joseph.p:2006,choi.ch:2006}. 
Equation~(\ref{isotropic}) with $b\kappa \gg  1$ suggests that a massive amplification of EO flow can be achieved over SH surfaces, but the controlled generation of such flows is by no means obvious, since both the slip length and the electric charge distribution on a SH surface are inhomogeneous and often anisotropic. Despite its fundamental and practical significance EO flow over SH surfaces has received little attention. Recently, \cite{Squires08} investigated EO flow past inhomogeneously charged, flat slipping surfaces in the case of thick channels ($h \gg L$) and  thin EDL ($\lambda_D \ll L$) and predicted negligible flow enhancement in case of an uncharged liquid-gas interface, which has been confirmed by molecular dynamics simulations~\citep{Huang08}.
However, this work cannot be trivially extended to the general case of thick EDL ($\lambda_D \gg L$), where improved efficiency of electrokinetic energy conversion is expected~\citep{heyden2006}. For thick EDL, we might also expect anisotropic EO flows transverse to the applied electric field, as in the case of rough, no-slip charged surfaces \citep{ajdari.a:2001}.

In this paper we provide analytical solutions for electro-osmotic flows over weakly charged, textured slipping surfaces. We show that the  electro-osmotic mobility is generally a tensorial property of the surface, reflecting nonlinear averaging of the slip-length and charge profiles, and is not trivially related to the hydrodynamic slip tensor \citep{Bazant08}.  In~\S\,\ref{sec:prob-def}, we give basic principles, formulate the problem and obtain general solutions for longitudinal and transverse textures. Effective slip lengths and exact solutions for EO velocity over stick-slip stripes modeling SH surfaces are derived in~\S\,\ref{sec:sup-hyd}. Implications for the use of SH  surfaces to enhance EO flows are discussed in~\S\,\ref{sec:discussion}, followed by concluding remarks in~\S\,\ref{sec:conclusion}.

\section{General theory}\label{sec:prob-def}

\subsection{ Interfacial mobility tensor }

\cite{ajdari.a:2001} pointed out that linear electrokinetic phenomena are generally tensorial in space and showed that microchannels with both charge and height variations can exhibit transverse electrokinetic effects. Here, we ascribe analogous behavior to a `thin interface', whose thickness $\lambda$ is much smaller than the geometrical scale $h$, by defining a tensorial electro-osmotic mobility via ${\bf u}_s = {\bf M}_e {\bf E}$, where ${\bf E}$ is the electric field  and ${\bf u}_s$ is the effective fluid slip velocity  just outside the interface (relative to the surface velocity).  This is analogous to the tensorial hydrodynamic mobility ${\bf M}_h$, defined by ${\bf u}_s = {\bf M}_h {\mbox{\boldmath{$\tau$}}}$ in terms of the normal traction ${\mbox{\boldmath{$\tau$}}}= {\mbox{\boldmath{$\hat{n}\cdot\sigma$}}}$,  or equivalently, to the slip-length tensor ${\bf b}={\bf M}_h\eta$ defined by  ${\bf u}_s = {\bf b} {\mbox{\boldmath{$\dot{\gamma}$}}}$ for a Newtonian fluid, where  ${\mbox{\boldmath{$\dot{\gamma}$}}}={\mbox{\boldmath{$\tau$}}}/\eta$ is the strain rate \citep{Bazant08}. We combine these effects in a general `interfacial constitutive relation'
\begin{equation}
\left( \begin{array}{c} {\bf u}_s \\ {\bf j}_s \end{array} \right) = \left( \begin{array}{cc} {\bf M}_h & {\bf M}_e \\ {\bf M}_e & {\bf K}_s \end{array} \right) 
\left( \begin{array}{c} {\mbox{\boldmath{$\tau$}}} \\ {\bf E} \end{array} \right) 
= {\mbox{\boldmath{${\cal M}$}}}\,  \left( \begin{array}{c} {\mbox{\boldmath{$\tau$}}} \\ {\bf E} \end{array} \right),   \label{mobility}
\end{equation}
which acts as an effective boundary condition on the quasi-neutral bulk fluid, 
where  ${\bf K}_s$ is a tensorial surface conductivity and ${\bf j}_s$ is the surface current density (in excess of the extrapolated bulk current density, integrated over the interface).  Using matched asymptotic expansions \citep{Chu07}, Eq. (\ref{mobility}) can be derived by considering a semi-infinite quasi-equilibrium electrolyte and  solving for the velocity ${\bf u}_s$  and current ${\bf j}_s$ `at infinity' relative to the interfacial thickness, e.g. $\lambda = \max\{ \lambda_D, L \}$ for a periodic texture of period $L$ or a (non-fractal) random texture with correlation length $L$.

Following \cite{Bazant08}, we note some basic physical constraints on ${\mbox{\boldmath{${\cal M}$}}}$. In most cases, we expect ${\mbox{\boldmath{${\cal M}$}}}$ to be symmetric, as assumed in (\ref{mobility}), by analogy with Onsager's relations for bulk non-equilibrium thermodynamics~\citep{degroot_book}. Indeed, this hypothesis can be rigorously established for Stokes flows over a broad class of patterned surfaces~\citep{kamrin09}. Here, we focus on the electro-osmotic mobility ${\bf M}_e$ for patterned slipping surfaces by calculating the anisotropic electro-osmotic flow in response to an applied electric field, but according to (\ref{mobility}) the same tensor also provides the `streaming surface current' ${\bf j}_s= {\bf M}_e {\mbox{\boldmath{$\tau$}}}$ in response to an applied shear stress. 

The interfacial mobility ${\mbox{\boldmath{${\cal M}$}}}$ must be positive definite for a passive interface, which produces entropy and does not do work on the fluid.   In general, the second-rank tensors ${\bf M}_h$, ${\bf M}_e$ and ${\bf K}_s$ could be represented by $3\times 3$ matrices to allow for normal flux of fluid (or charge) into a porous (or conducting) surface, driven by normal electric fields (or tensile stresses), but here we will only consider $2\times 2$ matrices in the coordinates of the tangent plane to describe  impermeable, insulating surfaces. This simplifies our analysis, since any symmetric, positive definite $2\times 2$ matrix is diagonalized by a rotation: 
\begin{equation}
{\bf M}_e= {\bf S}_{\theta} \left(
\begin{array}{cc}
 M_e^{\parallel} & 0\\
 0 & M_e^{\perp}
\end{array}
\right) {\bf S}_{-\theta} ,
\qquad
{\bf S}_{\theta}=
\left(
\begin{array}{cc}
 \Xcos~\theta &  \Xsin~\theta \\
 -\Xsin~\theta & \Xcos~\theta
\end{array}
\right). \label{Tensor_Rotation}
\end{equation}
Once the orthogonal eigen-directions $\theta=0, \pi/2$ are identified, the problem reduces to computing the two eigenvalues, $M_e^{\parallel}$ and $M_e^\perp$, which attain the maximal and minimal directional mobilities, respectively. 

\subsection{ Weakly charged, nano-scale striped patterns }

\begin{figure}
\begin{minipage}[b]{0.5\linewidth}
\centering
\includegraphics[height=5cm]{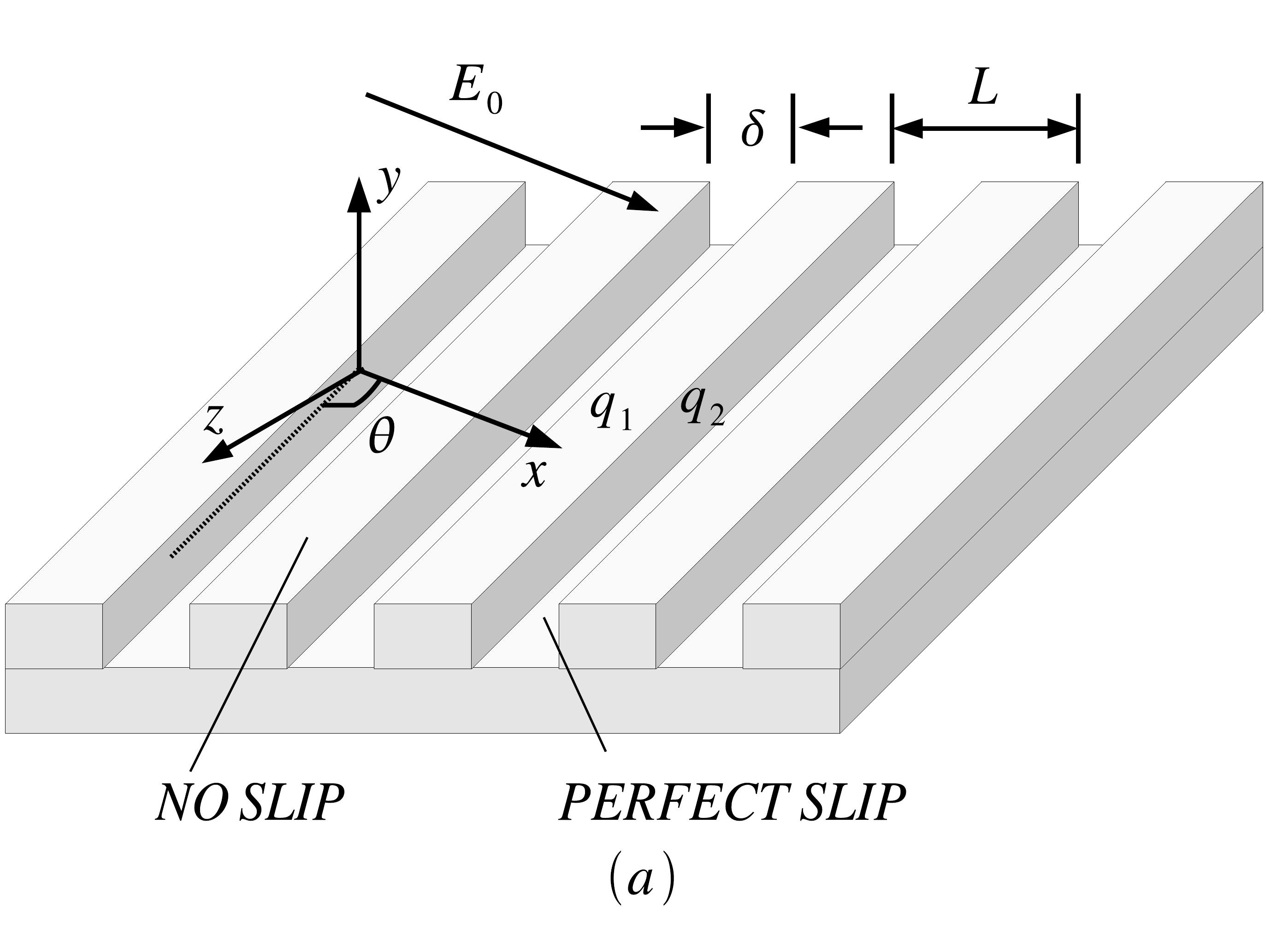}
\end{minipage}
\hspace{0.5cm}
\begin{minipage}[b]{0.5\linewidth}
\centering
\includegraphics[height=5cm]{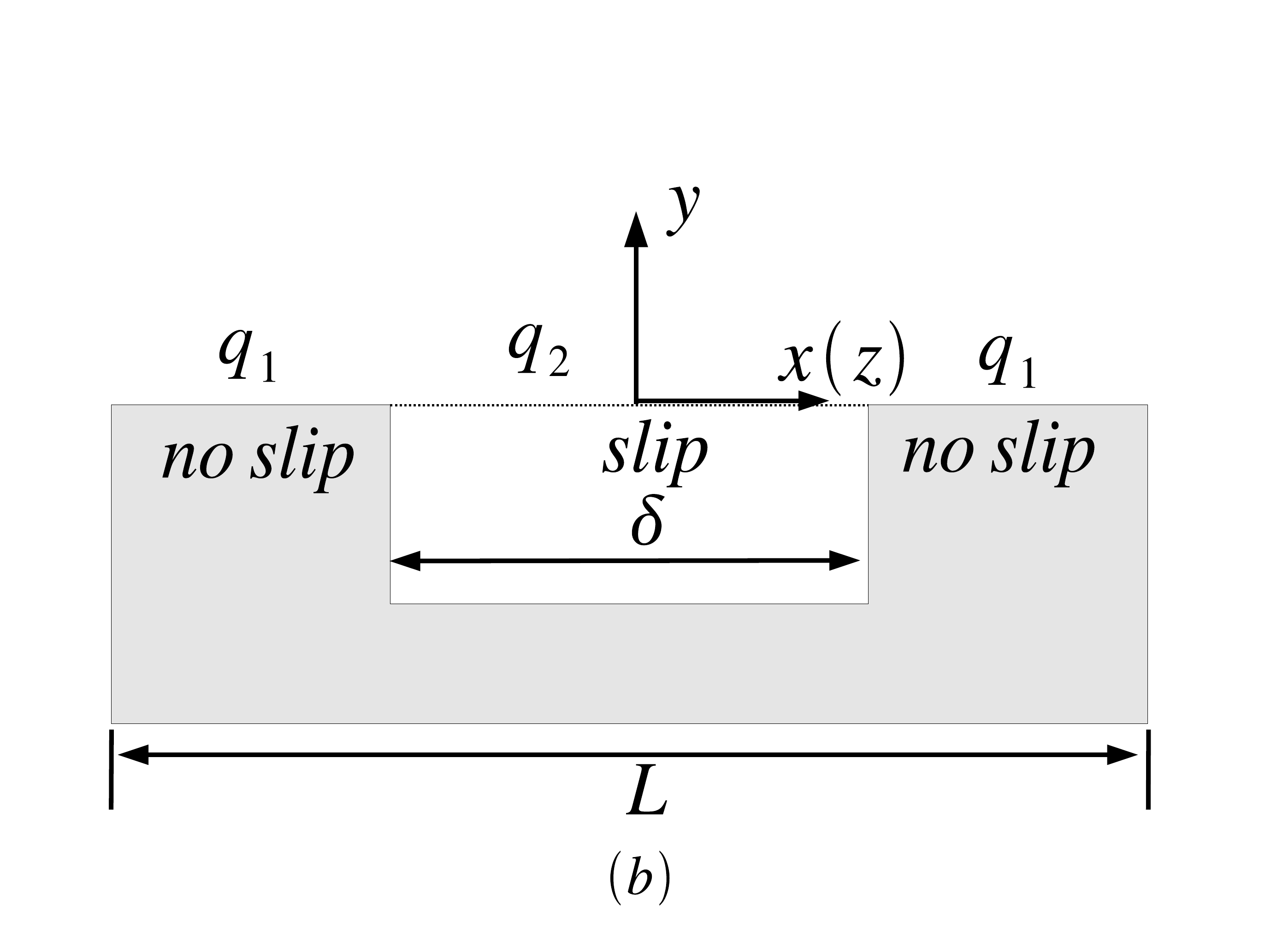}
\end{minipage}
\caption{(a) Sketch of SH stripes: $\theta=\pi/2$ corresponds to transverse, whereas $\theta=0$ to longitudinal stripes; (b) Situation in (a) is approximated by a periodic cell of size $L$, with equivalent flow boundary conditions on gas/liquid and solid/liquid interface. }\label{fig:geometry}
\end{figure}

To highlight effects of anisotropy, we focus on flat patterned SH surfaces consisting of periodic stripes, where the surface charge density $q$ and local (scalar) slip length $b$ vary only in one direction. In the case of thin channels ($h \ll L$), striped surfaces provide rigorous upper and lower bounds on the effective slip (eigenvalues of ${\bf M}_h$) over all possible two-phase patterns~\citep{feuillebois.f:2009}; for thick channels, sinusoidal stripes also bound the effective slip for arbitrary perturbations in surface height and/or slip length~\cite{kamrin09}. Striped SH surfaces have also been used for reduction in pressure-driven flows~\citep{ou2005}, with a typical geometry sketched in  Fig.~\ref{fig:geometry}a corresponding to Cassie's state of a roughly flat liquid surface over gas bubbles trapped in wells. By symmetry, the eigen-directions of ${\bf M}_e$, ${\bf M}_h$, and ${\bf K}_s$ for a striped surface correspond to longitudinal ($\theta=0$) and transverse ($\theta=\pi/2$) alignment with the applied electric field or shear stress, so we need only compute the eigenvalues for these cases using (\ref{Tensor_Rotation}).

We consider a semi-infinite electrolyte in the region $y>0$ above a flat patterned surface of period $L$ at $y=0$ subject to an electric field $E_0$ in the $x$ direction. The electrostatic potential is given by $\phi(x,y,z)=-E_0x + \psi(x,y,z)$, where $\psi$, is the perturbation due to diffuse charge. For nano-scale patterns ($L < 1\mu$m), we can neglect convection ($Pe = \langle q\rangle E_0L/\eta\kappa D \ll 1$ for a typical ionic diffusivity $D$), so that  $\psi(x,y,z)$ is independent of the fluid flow. We also assume weak fields $|E_o| L \ll |\psi|$ and weakly charged surfaces ($|\psi| \ll kT/ze=25/z$ mV at room temperature) for a  $z:z$ electrolyte, so that $\psi$ satisfies the Debye-H\"uckel equation with a boundary condition of prescribed surface charge,
\begin{equation}
\Delta \psi = \frac{\partial^2 \psi}{\partial x^2} + \frac{\partial^2 \psi}{\partial y^2} + \frac{\partial^2 \psi}{\partial z^2}= \kappa^2\psi, 
\ \ \ \ -\varepsilon \frac{\partial \psi}{\partial y}(x,0,z) = q(x,z),
\label{linear-PB}
\end{equation}
where, $\kappa={\lambda_D}^{-1}=\displaystyle ({2z^2 e^2 n_{\infty}}/{\varepsilon KT})^{1/2}$, is the
inverse screening length. 
In this limit, we can neglect surface conduction (which tends to reduce electro-osmotic flow) compared to bulk conduction ($Du = |{\bf j}_s|/\sigma E_0 L \ll 1$), so we will only discuss the tensors ${\bf M}_e$ and ${\bf M}_h$. 

For transverse stripes, $q=q(x)=q(x+L)$, we expand $q(x)$ in a Fourier series,
\begin{equation}
q(x)=\langle q \rangle+\sum_{n=1}^{\infty}(A_n \Xsin(\lambda_n x) + B_n \Xcos(\lambda_n x)),  \label{q}
\end{equation}
where $\langle q \rangle$ is the mean surface charge, and solve (\ref{linear-PB})  by separation of variables,
\begin{equation}
\psi=\psi(x,y)=\frac{\langle q \rangle}{\varepsilon\kappa} e^{-\kappa y}+\sum_{n=1}^{\infty} \frac{1}{\varepsilon\sqrt{\kappa^2+\lambda_n^2}}(A_n \Xsin(\lambda_n x) + B_n \Xcos(\lambda_n x))e^{-\sqrt{\kappa^2+\lambda_n^2} y}, \label{potential-sol}
\end{equation}
where $\lambda_n=2n\pi/L$. For longitudinal stripes, $q=q(z)=q(z+L)$, the potential $\psi=\psi(y,z)$ has exactly the same form  (\ref{potential-sol}) with $x$ replaced by $z$.

The fluid flow satisfies Stokes' equations with an electrostatic body force,
\refstepcounter{equation}
$$
\eta \Delta {\bf u}= -\varepsilon \Delta \psi \nabla \phi + \nabla p, \label{Stokes} \quad
\bnabla\bcdot{\bf u}=0. \label{continuity}   \eqno{(\theequation{\mathit{a},\mathit{b}})}
$$
To describe local hydrodynamic slip, we apply Navier's boundary condition
\begin{equation}
{\bf u}(x,0,z) = b(x,z) \frac{\partial {\bf u}}{\partial y}(x,0,z), \ \ \ \ 
\hat{{\bf y}}\cdot {\bf u}(x,0,z) = 0.  \label{slip-BC}
\end{equation}
Far from the surface, $\bf u$ approaches the effective, electro-osmotic `slip' velocity, \\$\lim_{y\to\infty}{\bf u}(x,y,z)={\bf u}_0$, and the derivatives of ${\bf u}$ remain bounded. By definition, the flow is two-dimensional in the eigen-directions of the surface. 

For transverse stripes, we have ${\bf u}=(u(x,y), v(x,y), 0)$,
$u(x,0)=b(x)u_y(x,0)$, and $v(x,0)=0$.  Assuming constant $\eta$ and $\varepsilon$ and using (\ref{linear-PB}), we can write (\ref{Stokes}) as,
\begin{equation}
\Delta {\bf u}(x,y) = -\frac{\varepsilon \kappa^2}{\eta}\psi(x,y)\nabla\phi(x,y) + \frac{1}{\eta}\nabla p, \label{Stokes-modified}
\end{equation}
where the pressure can be eliminated by taking the curl of both sides,
\[
\Delta \big(\nabla \times {\bf u}\big) = -\frac{\varepsilon \kappa^2}{\eta}(E_0 \hat{\bf x}\times \nabla \psi).
\]
Taking another curl of this equation and using incompressibility we obtain
\begin{equation}
\nabla^4 u= \frac{\varepsilon \kappa^2 E_0}{\eta}\Big(\frac{\partial^2 \psi}{\partial y^2}+\frac{\partial^2 \psi}{\partial z^2}\Big). \label{curl-Stokes-u}
\end{equation}
where  $\nabla^4 \psi=\kappa^4\psi$ from (\ref{linear-PB}). The general solution of (\ref{curl-Stokes-u}) for $u(x,y)$ has the form,
\begin{equation}
u(x,y)=u_0 + \sum_{n=1}^{\infty}\big(P_n \Xsin(\lambda_n x) + Q_n \Xcos(\lambda_n x) \big)e^{-\lambda_n y} + \frac{\varepsilon  E_0}{\kappa^2\eta}\frac{\partial^2 \psi}{\partial y^2}, \label{solution-u}
\end{equation}
where $P_n$, $Q_n$ are unknown coefficients and the last term involving $\psi(x,y)$, can be obtained from~(\ref{potential-sol}).
The slip boundary condition then determines the coefficients,  $\{P_n, Q_n\}$, and the electro-osmotic slip, $u_0=M_e^\perp E_o$.

For longitudinal stripes, the flow is also two dimensional: ${\bf u}=(0, v(y,z), u(y,z))$, 
$u(0,z)=b(z)u_y(0,z)$, and $v(0,z)=0$, where $u$ is again the tangential velocity. Similar steps lead to ~(\ref{curl-Stokes-u}) for $u(y,z)$, using~(\ref{linear-PB}) with $\psi=\psi(y,z)$. The general solution now takes the form,
\begin{equation}
u(y,z)=u_0 + \sum_{n=1}^{\infty}\big(P_n \Xsin(\lambda_n z) + Q_n \Xcos(\lambda_n z) \big)e^{-\lambda_n y} + \frac{\varepsilon  E_0}{\eta}\psi. \label{solution-u-long}
\end{equation} 
where $\{P_n, Q_n\}$ and $u_0=M_e^\| E_o$ are determined by the slip boundary condition.

\section{Striped super-hydrophobic surfaces} \label{sec:sup-hyd}

\subsection{ Hydrodynamic mobility tensor }

To illustrate the theory, we consider an idealized, flat, periodic, charged, striped SH surface in the Cassie state, sketched in figure~\ref{fig:geometry}a, where  the liquid-solid interface has no slip ($b_1=0$) and the liquid-gas interface has perfect slip ($b_2=\infty$). 
Let $\phi_1$ and $\phi_2=\delta/L$ be the area fractions of the solid and gas phases with $\phi_1+\phi_2=1$. Our results apply to a single surface in a thick channel ($h\gg \max\{ \lambda_D, L \}$), where effective hydrodynamic slip is determined by flow at the scale of roughness~\citep{bocquet2007}, but not to thin channels ($h \ll \min\{ \lambda_D, L \}$) where the effective slip scales with the channel width~\citep{feuillebois.f:2009}. 

Pressure-driven flow past stick-slip stripes has been analyzed  and shown to
depend on the direction of the flow \citep{Lauga03,cottin.c:2004,sbragaglia.m:2007}. 
Following ~\cite{Bazant08}, the hydrodynamic slip tensor ${\bf M}_h$ must have the form (\ref{Tensor_Rotation}), where the eigenvalues are 
\begin{equation}\label{effectiveslip}
M_h^\perp = \frac{b_{\rm eff}}{\eta} = \frac{L}{2 \pi \eta}\Xlog\Big[\Xsec\left(\frac{\pi \phi_2 }{2} \right)\Big]
\ \ \ \mbox{ and } \ \ \ M_h^\| = 2 M_h^\perp.
\end{equation}
Note that the effective slip for parallel stripes $M_h^\|$ is twice that of perpendicular stripes, $M_h^\|$, analogous to the result from slender body theory that an vertically oriented elongated body sedimenting due to its own weight falls twice faster then if it were oriented horizontally (\cite{Batchelor70}). 

%

\subsection{Electro-osmotic mobility tensor }

Using (\ref{Tensor_Rotation}), we need only calculate the eigenvalues of ${\bf M}_e$ for transverse and longitudinal stripes. 
For our model SH surface with transverse stripes ($\theta=0$), the region $|x|\leq \thalf \delta$ has $b=\infty$ (i.e. $u_y(x,0)=0$) and $q=q_1$, while the region $ \thalf \delta < |x|\leq  \thalf L$ has 
$b=0$ and $q=q_2$.
Imposing these boundary conditions on the general solution~(\ref{solution-u}) yields a dual cosine series,
\begin{subeqnarray}
\frac{\langle q \rangle E_0}{\eta} + \sum_{n=1}^{\infty}\Big(\lambda_n Q_n + \frac{\gamma_n^2 E_0 B_n}{\eta \kappa^2} \Big)\Xcos(\lambda_n x) =0,\quad  \forall~|x|\leq  \thalf\delta, \label{slip-dual-series}\\
u_0 + \frac{\langle q \rangle E_0}{\eta\kappa} + \sum_{n=1}^{\infty}\Big(Q_n + \frac{\gamma_n E_0 B_n}{\eta \kappa^2} \Big)\Xcos(\lambda_n x) =0,\quad  \forall~ \thalf\delta <|x|\leq  \thalf L, \label{noslip-dual-series}
\end{subeqnarray}
where $\gamma_n=\sqrt{\lambda_n^2+\kappa^2}$. (The sine terms vanish due to symmetry.)

\begin{center}
\begin{figure}
\hspace{1cm}
 \centerline{\includegraphics[height=5cm,width=13.5cm]{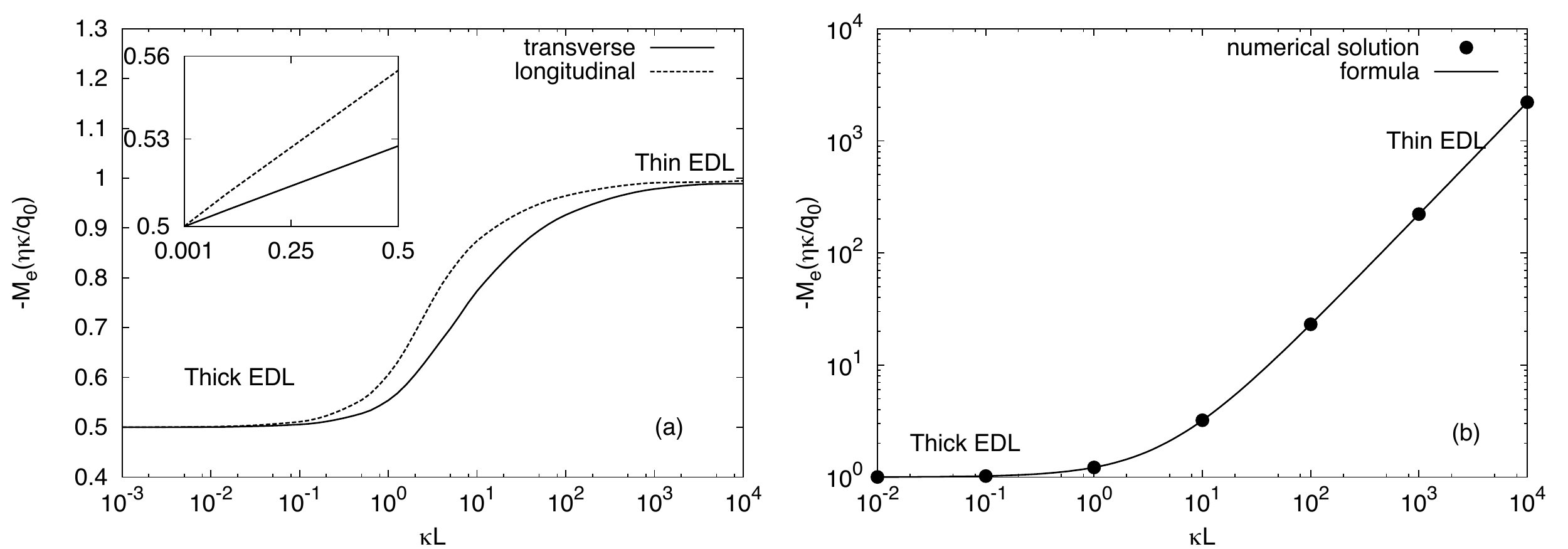}}
  \caption{Eigenvalues of the electro-osmotic slip tensor ${\bf M}_e$ for stick-slip stripes of period $L$ and slipping area fraction $\delta/L=1/2$ as a function of electric double-layer (EDL) thickness $\kappa^{-1}$. Numerical solutions are compared with exact results in the text for the limits $\kappa L \to 0, \infty$ for (a) an uncharged slipping interface $q_1=q_0, q_2=0$, with thick EDL limit clarified in the inset, and (b) a surface of constant charge $q_1=q_2=q_0$.
}\label{fig:results}
\end{figure}
 \end{center}

For the general case $q_1\neq q_2$, the  dual series can be solved numerically for $M_e^\perp = u_0/E_0$ (see Fig. ~\ref{fig:results}) by truncating the series and taking the inner products with $1$ and $\cos(\lambda_n x)$, but exact results are possible in the thin and thick EDL limits. (Below, we also give an exact solution for any value of $\kappa L$ in the case $q_1=q_2$.)  In the thin EDL limit, $\lambda_n/\kappa \to 0$, we have $\gamma_n/\kappa \to 1$.
Since $B_n$ are Fourier cosine coefficients of $q(x)$,
the dual series can be written as,
\begin{subeqnarray}
\sum_{n=1}^{\infty}\lambda_n Q_n \Xcos(\lambda_n x) =-\frac{q_2 E_0}{\eta},\quad  \forall~|x|\leq  \thalf \delta, \\
\Big(u_0 + \frac{q_{1} E_0}{\eta\kappa}\Big)+ \sum_{n=1}^{\infty}Q_n \Xcos(\lambda_n x) =0,\quad  \forall~ \thalf\delta < |x|\leq  \thalf L.\label{dual-trans-thin}
\end{subeqnarray}
This dual series can be solved exactly~\citep{sneddon} to obtain
\begin{equation}
M_e^{\perp, thin} = \frac{u_0}{E_0} =-\frac{q_1 +  2 q_2\kappa    b_{\rm eff} }{\eta\kappa}.
\label{sol-thin-trans}
\end{equation}
In the thick EDL limit, $\lambda_n/\kappa \gg1$, the dual series ~(\ref{slip-dual-series}) takes the form,
\begin{subeqnarray}
\sum_{n=1}^{\infty}\lambda_n\Big(Q_n + \frac{\gamma_n E_0 B_n}{\eta \kappa^2} \Big)\Xcos(\lambda_n x) = -\frac{E_0}{2\eta}(\langle q \rangle + q_2)~~\forall~|x|\leq  \thalf \delta, \label{slip-dual-series-3}\\
u_0 + \frac{\langle q \rangle E_0}{\eta\kappa} + \sum_{n=1}^{\infty}\Big(Q_n + \frac{\gamma_n E_0 B_n}{\eta \kappa^2} \Big)\Xcos(\lambda_n x) =0,\quad  \forall~ \thalf \delta < |x|\leq  \thalf L,
\end{subeqnarray}
which can again be solved exactly to obtain the thick-DL electro-osmotic mobility,
\begin{equation}
M_e^{\perp, thick} = \frac{ u_0}{E_0} = -\frac{\langle q \rangle}{\eta\kappa} \left[  1 +  \left(\frac{ \langle q \rangle+q_2}{\langle q \rangle}\right)  b_{\rm eff} \kappa \right].
\label{sol-thick-trans}
\end{equation}

Rotating the SH surface by $\theta=\pi/2$ for longitudinal stripes, the region $|z|\leq \thalf \delta$ has $b=\infty$ and $q=q_1$, while the region $ \thalf \delta < |z|\leq  \thalf L$ has $b=0$ and $q=q_2$.
Applying boundary conditions to (\ref{solution-u-long}), we obtain another dual cosine series
\begin{subeqnarray}
\frac{\langle q \rangle E_0}{\eta} + \sum_{n=1}^{\infty}\Big(\lambda_n Q_n + \frac{E_0 B_n}{\eta} \Big)\Xcos(\lambda_n z) =0,\quad  \forall~|z|\leq  \thalf \delta, \label{slip-dual-series-long}\\
u_0 + \frac{\langle q \rangle E_0}{\eta\kappa} + \sum_{n=1}^{\infty}\Big(Q_n + \frac{E_0 B_n}{\eta \gamma_n } \Big)\Xcos(\lambda_n z) =0,\quad  \forall~ \thalf \delta <|z|\leq  \thalf L, \label{noslip-dual-series-long}
\end{subeqnarray}
which can be solved numerically (Fig. 2) or exactly for thin and thick EDL. 
In the thin EDL limit, $\gamma_n/\kappa \to 1$, the dual series~(\ref{slip-dual-series-long}) can be simplified using the $q(x)$ series to obtain again~(\ref{dual-trans-thin}) and thus 
\begin{equation}
M_e^{\|, thin} = M_e^{\perp, thin} = M_e^{thin} =-\frac{q_1 +  2 q_2\kappa    b_{\rm eff} }{\eta\kappa}.  \label{sol-thin-long}
\end{equation}
Therefore, we find that the electro-osmotic mobility tensor is isotropic in the thin DL limit, ${\bf M}_e^{thin} = M_e^{thin} {\bf I}$, consistent with the examples of \cite{Squires08}. In general, ${\bf M}_e$ must be isotropic for any flat patterned surface in the thin DL limit, since the effective EO slip velocity is equal to the surface averaged EO slip velocity \citep{ramos03}, and thus always in the direction of ${\bf E}_0$. 

In the thick EDL limit,
we have $\lambda_n/\gamma_n \to 1$, $\lambda_n/\kappa \gg1$, and the dual series reduces to,
\begin{subeqnarray}
\sum_{n=1}^{\infty}\lambda_n\Big(Q_n + \frac{E_0 B_n}{\gamma_n\eta} \Big)\Xcos(\lambda_n z) = -\frac{\langle q\rangle E_0}{\eta},\quad  \forall~|z|\leq  \thalf \delta,\\
\Big(u_0 + \frac{\langle q \rangle E_0}{\eta\kappa}\Big) + \sum_{n=1}^{\infty}\Big(Q_n + \frac{E_0 B_n}{\eta \gamma_n } \Big)\Xcos(\lambda_n z) =0,\quad  \forall~ \thalf \delta <|z|\leq   \thalf L.
\end{subeqnarray}
which can be solved exactly 
\begin{equation}
M_e^{\|, thick} = -\frac{\langle q \rangle}{\eta\kappa}\left( 1 +  2\kappa b_{\rm eff}\right).  \label{sol-thick-long}
\end{equation}
Since $M_e^{\perp, thick}\neq M_e^{\|, thick}$, we see that ${\bf M}_e$ becomes anisotropic for thick DL.

\section{ Discussion }\label{sec:discussion} 

\subsection{Uncharged liquid-gas interface }

It is instructive to set the charge to zero on the slipping surface, $q_2=0$, to describe the Cassie state of a SH surface with an uncharged liquid-gas interface.   The EO flow is then related only to the charge $q_1=q_0$ on the no-slip liquid-solid interface. First we consider the thin EDL limit, where the EO mobility is generally isotropic, as noted above.
Using~(\ref{sol-thin-long}) we obtain the simple result of \cite{Squires08},
\begin{equation}
{\bf M}_e^{thin} = M_e^{thin} {\bf I}, \ \ \ \mbox{where } \ \ \ M_e^{thin}=-\frac{q_0}{\eta\kappa} = -\frac{\langle q \rangle}{\phi_1\eta\kappa},
\label{cassie-formula-long}
\end{equation}
where the EO mobility is the same as for a homogeneous, solid no-slip surface with charge density $q_0$, regardless of the orientation or area fraction of the slipping stripes. In other words, there is no EO flow enhancement due to the slipping regions.  As explained by \cite{Squires08}, the liquid appears to slip on the charged liquid-solid interface by electro-osmosis, but without any retarding shear stress or amplifying electro-osmotic flow on the uncharged, perfectly slipping liquid-gas interface.

The flow is anisotropic for any finite EDL thickness, and in this case, there is a simple relationship between the electro-osmotic and hydrodynamic mobility tensors in the thick EDL limit. Using~(\ref{sol-thick-trans}) and ~(\ref{sol-thick-long}) for the EO mobility eigenvalues and comparing with (\ref{effectiveslip}), we find 
\begin{equation}
{\bf M}_e^{thick} = - \langle q \rangle \left( \frac{{\bf I}}{\eta \kappa} + {\bf M}_h \right) =
- \frac{ \langle q \rangle }{\eta \kappa} \left( {\bf I} + \kappa {\bf b} \right),   \label{uncharged-thick}
\end{equation}
which is a natural tensorial generalization of the classical formula (\ref{isotropic}).  In the limit $\kappa L \to 0$ (similar to the thick channel limit), hydrodynamic slip becomes negligible, and the EO flow is isotropic and driven by the average surface charge, $M_e^{thick} \to -  \langle q \rangle / \eta \kappa$.

%

From the thick EDL results of this case we see that the EO velocities are even smaller than in case of thin EDL. Even if ~(\ref{cassie-formula-long}) is used for $\kappa L\approx 1$, the EO mobility of the SH surface is still smaller than having a constant charge on the surface without any slipping regions. This is consistent with the  molecular dynamics simulations of ~\cite{Huang08} for perpendicular stripes for 1~M NaCl solution confined in parallel walls with no charge on the liquid-gas interface, which also showed no enhancement of EO flow.  Since the effective slip length depends on the pattern, however, some flow enhancement may be possible for thick EDL, even with an uncharged liquid-gas interface, provided that effective slip length is quite large.

\subsection{Charged liquid-gas interface}

The situation is very different if the liquid-gas interface carries some charge. There is  strong experimental evidence that the air-water interface is negatively charged due presence of excess OH$^-$ ions ~(\cite{Takahashi05}) supported by the molecular dynamics simulations of~\cite{Zangi05}. Adsorption of ions on the water-air interface has also been justified theoretically by \cite{Shchekin05}. 

To gain some analytical insight into the possible EO flow enhancement, we consider the special case of uniform surface charge $q_1=q_2=q_0$. In this case, an exact solution is possible for arbitrary EDL thickness and any stripe orientation:
\begin{equation}
{\bf M}_e = M_e {\bf I}, \ \ \ \mbox{ where}  \ \ \ 
M_e = - q_0 \left( \frac{1}{\eta\kappa}  + M_h^\| \right) = -\frac{q_0}{\eta\kappa}(1+ 2 b_{\rm eff}\kappa). 
\label{const-q-formula}
\end{equation}
Clearly, a large EO flow enhancement, similar to that of an isotropic surface (\ref{isotropic}), is possible with a charged liquid-gas interface (Fig.~\ref{fig:results}b). 
This conclusion holds even if the charge is not homogeneous. For thin EDL $\kappa L\sim 10^{3}$, and $q_2/q_1\sim 0.1-1$, $\phi_2\sim 0.5$, the theory predicts EO flow enhancement by factor of $10-100$. For thick EDL, this factor approaches $1$, but is offset by the prefactor $\kappa^{-1}$ in the case of constant charge. We conclude that in practical applications of electrokinetic phenomena involving SH  surfaces, e.g. for microfluidic pumping or energy conversion, electrokinetic enhancement is possible with a charged liquid-gas interface, perhaps by an order of magnitude, over a homogeneous no-slip surface.

\section{Conclusion}\label{sec:conclusion}

We have analyzed electro-osmotic flows over patterned surfaces of non-uniform charge and local slip length. Unlike the approach of~\cite{Squires08}, we have obtained general solutions for arbitrary EDL thickness, surface charge distribution and slip variation, which are qualitatively different from the thin EDL limit.  We have shown that the electrokinetic response of a patterned slipping surface is generally anisotropic and describable by a third rank interfacial mobility tensor (\ref{mobility}), analogous to the tensorial linear response of a thin microchannel \citep{ajdari.a:2001}. For stick-slip stripes, we have calculated the electro-osmotic mobility tensor and shown that it is not simply related to the hydrodynamic mobility tensor \citep{Bazant08}. In particular, a tensorial generalization  (\ref{uncharged-thick}) of the classical formula (\ref{isotropic}) does not hold, except for uncharged slipping regions.

Our results provide some guidance for the design of SH surfaces for electrokinetic applications.  For an uncharged liquid-gas interface with thick EDL, our results are closely related to those of \cite{Lauga03} for pressure-driven flows, since EO flow enhancement is directly set by the hydrodynamic mobility tensor.  For an uncharged liquid-gas interface with thin EDL, we also confirm the result of \cite{Squires08} that there is no enhancement of  EO flow, but the general response for a SH surface can be quite different.  Our main conclusion is that a charged liquid-gas interface is required to achieve significant enhancement of EO flow. For a uniformly charged SH surface, we show that the EO mobility (\ref{const-q-formula}) is isotropic and can exhibit large enhancement from hydrodynamic slip, possibly by an order of magnitude.

This research was supported in part by the National Science Foundation, under
Contract DMS-0707641 (MZB).

\bibliographystyle{jfm}
\bibliography{jfm_supreet}

\begin{thebibliography}{37}
\expandafter\ifx\csname natexlab\endcsname\relax\def\natexlab#1{#1}\fi

\bibitem[Ajdari(2001)]{ajdari.a:2001}
{\sc Ajdari, A.} 2001 Transverse electrokinetic and microfluidic effects in
  micropatterned channels: Lubrication analysis for slab geometries. {\em Phys.
  Rev. E\/} {\bf 65}, 016301.

\bibitem[Ajdari \& Bocquet(2006)]{ajdari.a:2006}
{\sc Ajdari, A. \& Bocquet, L.} 2006 Giant amplification of interfacially
  driven transport by hydrodynamic slip: diffusio-osmosis and beyond. {\em
  Phys. Rev. Lett.\/} {\bf 96}, 186102.

\bibitem[Batchelor(1970)]{Batchelor70}
{\sc Batchelor, G.~K.} 1970 Slender-body theory for particles of arbitary
  cross-section in stokes flow. {\em J.~Fluid Mech.\/} {\bf 44}, 419--440.

\bibitem[Bazant \& Vinogradova(2008)]{Bazant08}
{\sc Bazant, M.~Z. \& Vinogradova, O.~I.} 2008 Tensorial hydrodynamic slip.
  {\em J.~Fluid Mech.\/} {\bf 613}, 125--134.

\bibitem[{Bocquet} \& Barrat(2007)]{bocquet2007}
{\sc {Bocquet}, L. \& Barrat, J.~L.} 2007 Flow boundary conditions from nano-
  to micro- scales. {\em Soft Matter\/} {\bf 3}, 685--693.

\bibitem[Choi {\em et~al.\/}(2006)Choi, Ulmanella, Kim, Ho \&
  Kim]{choi.ch:2006}
{\sc Choi, C.~H., Ulmanella, U., Kim, J., Ho, C.~M. \& Kim, C.~J.} 2006
  Effective slip and friction reduction in nanograted superhydrophobic
  microchannels. {\em Phys. Fluids\/} {\bf 18}, 087105.

\bibitem[Chu \& Bazant(2007)]{Chu07}
{\sc Chu, K.~T. \& Bazant, M.~Z.} 2007 Surface conservation laws at
  microscopically diffuse interfaces. {\em J. Colloid and Interface Science\/}
  {\bf 315}, 319--329.

\bibitem[Cottin-Bizonne {\em et~al.\/}(2004)Cottin-Bizonne, Barentin, Charlaix,
  Bocquet \& Barrat]{cottin.c:2004}
{\sc Cottin-Bizonne, C., Barentin, C., Charlaix, E., Bocquet, L. \& Barrat,
  J.~L.} 2004 Dynamics of simple liquids at heterogeneous surfaces:
  Molecular-dynamic simulations and hydrodynamic description. {\em Eur. Phys.
  J. E\/} {\bf 15}, 427.

\bibitem[Cottin-Bizonne {\em et~al.\/}(2003)Cottin-Bizonne, Barrat, Bocquet \&
  Charlaix]{cottin_bizonne.c:2003.a}
{\sc Cottin-Bizonne, C., Barrat, J.~L., Bocquet, L. \& Charlaix, E.} 2003
  Low-friction flows of liquid at nanopatterned interfaces. {\em Nat. Mater.\/}
  {\bf 2}, 237--240.

\bibitem[Cottin-Bizonne {\em et~al.\/}(2005)Cottin-Bizonne, Cross, Steinberger
  \& Charlaix]{charlaix.e:2005}
{\sc Cottin-Bizonne, C., Cross, B., Steinberger, A. \& Charlaix, E.} 2005
  Boundary slip on smooth hydrophobic surfaces: Intrinsic effects and possible
  artifacts. {\em Phys. Rev. Lett.\/} {\bf 94}, 056102.

\bibitem[Feuillebois {\em et~al.\/}(2009)Feuillebois, Bazant \&
  Vinogradova]{feuillebois.f:2009}
{\sc Feuillebois, F., Bazant, M.~Z. \& Vinogradova, O.~I.} 2009 Effective slip
  over superhydrophobic surfaces in thin channels. {\em Phys. Rev. Lett.\/}
  {\bf 102}, 026001.

\bibitem[Groot \& Mazur(1962)]{degroot_book}
{\sc Groot, S.~R.~De \& Mazur, P.} 1962 {\em Non-equilibrium Thermodynamics\/}.
  New York, NY: Interscience Publishers, Inc.

\bibitem[van~der Heyden {\em et~al.\/}(2006)van~der Heyden, Bonthuis, Stein,
  Meyer \& Dekker]{heyden2006}
{\sc van~der Heyden, F. H.~J., Bonthuis, D.~J., Stein, D., Meyer, C. \& Dekker,
  C.} 2006 Electrokinetic energy conversion efficiency in nanofluidic channels.
  {\em Nano Letters\/} {\bf 6}, 2232--2237.

\bibitem[Huang {\em et~al.\/}(2008)Huang, Cottin-Bizzone, Ybert \&
  Bocquet]{Huang08}
{\sc Huang, D.~M., Cottin-Bizzone, C., Ybert, C. \& Bocquet, L.} 2008 Massive
  amplification of surface-induced transport at superhydrophobic surfaces. {\em
  Phys.~Fluids\/} {\bf 20}, 092105.

\bibitem[Joly {\em et~al.\/}(2006)Joly, Ybert \& Bocquet]{joly.l:2006}
{\sc Joly, L., Ybert, C. \& Bocquet, L.} 2006 Probing the nanohydrodynamics at
  liquid-solid interfaces using thermal motion. {\em Phys. Rev. Lett.\/} {\bf
  96}, 046101.

\bibitem[Joly {\em et~al.\/}(2004)Joly, Ybert, Trizac \& Bocquet]{joly2004}
{\sc Joly, L., Ybert, C., Trizac, E. \& Bocquet, L.} 2004 Hydrodynamics within
  the electric double layer on slipping surfaces. {\em Phys. Rev. Lett.\/} {\bf
  93}, 257805.

\bibitem[Joseph {\em et~al.\/}(2006)Joseph, Cottin-Bizonne, Beno\v{\i}, Ybert,
  Journet, Tabeling \& Bocquet]{joseph.p:2006}
{\sc Joseph, P., Cottin-Bizonne, C, Beno\v{\i}, J.~M., Ybert, C., Journet, C.,
  Tabeling, P. \& Bocquet, L.} 2006 Slippage of water past superhydrophobic
  carbon nanotube forests in microchannels. {\em Phys. Rev. Lett.\/} {\bf 97},
  156104.

\bibitem[Kamrin {\em et~al.\/}(2009)Kamrin, Bazant \& Stone]{kamrin09}
{\sc Kamrin, K., Bazant, M.~Z. \& Stone, H.~A.} 2009 In preparation.

\bibitem[Khair \& Squires(2009)]{khair.as:2009}
{\sc Khair, A.~S. \& Squires, T.~M.} 2009 The influence of hydrodynamic slip on
  the electrophoretic mobility of a spherical colloidal particle. {\em Phys.
  Fluids\/} {\bf 21}, 042001.

\bibitem[Lauga {\em et~al.\/}(2007)Lauga, Brenner \& Stone]{lauga2005}
{\sc Lauga, E., Brenner, M.~P. \& Stone, H.~A.} 2007 {\em Handbook of
  Experimental Fluid Dynamics\/}, chap.~19, pp. 1219--1240. NY: Springer.

\bibitem[Lauga \& Stone(2003)]{Lauga03}
{\sc Lauga, E. \& Stone, H.~A.} 2003 Effective slip in pressure-driven stokes
  flow. {\em J.~Fluid Mech.\/} {\bf 489}, 55--77.

\bibitem[Muller {\em et~al.\/}(1986)Muller, Sergeeva, Sobolev \&
  Churaev]{muller.vm:1986}
{\sc Muller, V.~M., Sergeeva, I.~P., Sobolev, V.~D. \& Churaev, N.~V.} 1986
  Boundary effects in the theory of electrokinatic phenomena. {\em Colloid J.
  USSR\/} {\bf 48}, 606--614.

\bibitem[{Ou} \& {Rothstein}(2005)]{ou2005}
{\sc {Ou}, J. \& {Rothstein}, J.~P.} 2005 Direct velocity measurements of the
  flow past drag-reducing ultrahydrophobic surfaces. {\em Physics of Fluids\/}
  {\bf 17}, 103606.

\bibitem[Ramos {\em et~al.\/}(2003)Ramos, Gonz\'alez, Castellanos, Green \&
  Morgan]{ramos03}
{\sc Ramos, A., Gonz\'alez, A., Castellanos, A., Green, N.~G. \& Morgan, H.}
  2003 Pumping of liquids with ac voltages applied to asymmetric pairs of
  microelectrodes. {\em Phys. Rev. E\/} {\bf 67}, 056302.

\bibitem[Sbragaglia \& Prosperetti(2007)]{sbragaglia.m:2007}
{\sc Sbragaglia, M. \& Prosperetti, A.} 2007 A note on the effective slip
  properties. {\em Phys. Fluids\/} {\bf 19}, 043603.

\bibitem[Shchekin \& Borisov(2005)]{Shchekin05}
{\sc Shchekin, A.~K. \& Borisov, V.~V.} 2005 Thermodynamics of nucleation on
  the particles of salts?strong electrolytes: The allowance for ion adsorption
  in the droplet surface layer. {\em Colloid J.\/} {\bf 67}, 774?787.

\bibitem[Sneddon(1966)]{sneddon}
{\sc Sneddon, I.~N.} 1966 In {\em Mixed boundary value problems in potential
  theory\/}. North-Holland.

\bibitem[Squires(2008)]{Squires08}
{\sc Squires, T.~M.} 2008 Electrokinetic flows over inhomogeneously slipping
  surfaces. {\em Phys.~Fluids\/} {\bf 20}, 092105.

\bibitem[Squires \& Quake(2005)]{squires2005}
{\sc Squires, T.~M. \& Quake, S.~R.} 2005 Microfluidics: Fluid physics at the
  nanoliter scale. {\em Reviews of Modern Physics\/} {\bf 77}, 977.

\bibitem[{Stone} {\em et~al.\/}(2004){Stone}, {Stroock} \& {Ajdari}]{stone2004}
{\sc {Stone}, H.~A., {Stroock}, A.~D. \& {Ajdari}, A.} 2004 Engineering flows
  in small devices. {\em Annual Review of Fluid Mechanics\/} {\bf 36},
  381--411.

\bibitem[Takahashi(2005)]{Takahashi05}
{\sc Takahashi, M.} 2005 $\zeta$ potential of microbubbles in aqueous
  solutions: electrical properites of the gas-water interface. {\em
  J.~Phys.~Chem.~B\/} {\bf 109}, 21858--21864.

\bibitem[Vinogradova(1995)]{vinogradova.oi:1995a}
{\sc Vinogradova, O.~I.} 1995 Drainage of a thin liquid film confined between
  hydrophobic surfaces. {\em Langmuir\/} {\bf 11}, 2213.

\bibitem[Vinogradova(1999)]{vinogradova1999}
{\sc Vinogradova, O.~I.} 1999 Slippage of water over hydrophobic surfaces. {\em
  Int. J. Miner. Proc.\/} {\bf 56}, 31--60.

\bibitem[Vinogradova {\em et~al.\/}(1995)Vinogradova, Bunkin, Churaev,
  Kiseleva, Lobeyev \& Ninham]{vinogradova.oi:1995b}
{\sc Vinogradova, O.~I., Bunkin, N.~F., Churaev, N.~V., Kiseleva, O.~A.,
  Lobeyev, A.~V. \& Ninham, B.~W.} 1995 Submicrocavity structure of water
  between hydrophobic and hydrophilic walls as revealed by optical cavitation.
  {\em J. Colloid Interface Sci.\/} {\bf 173}, 443--447.

\bibitem[Vinogradova {\em et~al.\/}(2009)Vinogradova, Koynov, Best \&
  Feuillebois]{vinogradova.oi:2009}
{\sc Vinogradova, O.~I., Koynov, K., Best, A. \& Feuillebois, F.} 2009 Direct
  measurements of hydrophobic slipage using double-focus fluorescence
  cross-correlation. {\em Phys. Rev. Lett.\/} {\bf 102}, 118302.

\bibitem[{Vinogradova} \& {Yakubov}(2003)]{vinogradova2003}
{\sc {Vinogradova}, O.~I. \& {Yakubov}, G.~E.} 2003 Dynamic effects on force
  measurements. 2. lubrication and the atomic force microscope. {\em
  Langmuir\/} {\bf 19}, 1227--1234.

\bibitem[Zangi \& Engberts(2005)]{Zangi05}
{\sc Zangi, R. \& Engberts, Jan B. F.~N.} 2005 Physisorption of hydroxide ions
  from aqueous solution to a hydrophobic surface. {\em J.~Am.~Chem.~Soc.\/}
  {\bf 127}, 2272--2276.

\end{thebibliography}

\end{document}